\documentclass[12pt, draftclsnofoot, onecolumn]{IEEEtran}
\usepackage[mathcal]{euscript}
\usepackage{amssymb}
\usepackage{amssymb}
\usepackage{amsmath,bm,amsthm}
\usepackage{mnsymbol}
\usepackage{graphicx}
\usepackage{psfrag}
\usepackage{epstopdf}
\usepackage{epsf,epsfig}
\usepackage{subfigure}
\usepackage{cite,color}
\usepackage{mathrsfs}
\usepackage{multirow}
\usepackage{makecell}

\usepackage{amsmath}   % From the American Mathematical Society
\renewcommand\IEEEkeywordsname{Index Terms}
\newcommand{\jj}{\mathrm{j}}
\begin{document}

\title{Bounds and Approximations for the Distribution of a Sum of Lognormal Random Variables}

%\author{Fredrik Berggren, \emph{Senior Member, IEEE}}% <-this % stops a space
%\author{Author}%

\author{Fredrik Berggren
\thanks{The author is with Huawei Technologies Sweden AB, Stockholm, Sweden. (e-mail: fredrik.b@huawei.com).}}

\maketitle

\begin{abstract}
A sum of lognormal random variables (RVs) appears in many problems of science and engineering. For example, it is invloved in computing the distribution of recevied signal and interference powers for radio channels subject to lognormal shadow fading. Its distribution has no closed-from expression and it is typically characterized by approximations, asymptotes or bounds. We give a novel upper bound on the cumulative distribution function (CDF) of a sum of $N$ lognormal RVs. The bound is derived from the tangential mean-arithmetic mean inequality. By using the tangential mean, our method replaces the sum of $N$ lognormal RVs with a product of $N$ shifted lognormal RVs. It is shown that the bound can be made arbitrarily close to the desired CDF, and thus it becomes more accurate than any other bound or approximation, as the shift approaches infinity. The bound is computed by numerical integration, for which we introduce the Mellin transform, which is applicable to products of RVs. At the left tail of the CDF, the bound can be expressed by a single Q-function. Moreover, we derive simple new approximations to the CDF, expressed as a product $N$ Q-functions, which are more accurate than the previous method of Farley. 
\end{abstract}
\begin{IEEEkeywords}
Bound, shadow fading, Mellin transform, sum of lognormal random variables, tangential mean. 
\end{IEEEkeywords}

\section{Introduction}
\IEEEPARstart{T}{he} sum of lognormal random variables (RVs) appears in various problems, e.g., in cellular radio communications where it is used to model the received co-channel interference power, or the received signal power from different diversity branches, subject to lognormal shadow fading. Based on the distribution, it is possible to determine the probability of link outage, i.e., that the signal power, or signal-to-interference ratio (SIR) falls below a given threshold. No closed-form expression has been presented for its cumulative distribution function (CDF) and plenty of research papers concerning the properties of the CDF have been published for this  long-standing problem. For example, some methods approximate the sum with another lognormal RV whose moments or cumulants are matched to those of the sum of RVs \cite{Beaulieu0}, \cite{Fenton},\cite{Schwarz}. Furthermore, moment matching by numerical integration of the moment generating function (MGF) \cite{Tellambura1} \cite{Mehta} or of the characteristic function (CF) \cite{Beaulieu1} has been proposed. Direct numerical integration of the Laplace transform \cite{Miles} has also been exploited. A good fit to the CDF by approximations based on other probability distributions has also been reported, e.g., \cite{Zhao},\cite{Nie},\cite{Liu},\cite{DiRenzo}, \cite{Li},\cite{Lam}.  

Another direction is to bound the CDF. In \cite{Slimane}, order statistics was used to derive upper- and lower bounds on the CDF for a sum of $N$ independent lognormal RVs, where the bounds were given either by Q-functions or on integral form. This methodology was extended in \cite{Tellambura} to bounds for the cases of  $N=2$ or $N=3$ arbitrarily correlated lognormal RVs, or  of $N$  equally correlated lognormal RVs, where the bounds were given on integral form. The geometric mean-arithmetic mean inequality was utilized in \cite{Berggren1} to obtain an upper bound on the CDF, described by a Q-function, for any $N$ and for RVs with arbitrary correlation. Lower bounds on the CDF have been given for the special case $N=2$  expressed by the Q-function \cite{Zhu},\cite{Xiao} or by the Marcum Q-function \cite{Beaulieu2}. In \cite{Berggren2}, a bound was given on the error between the CDF of an approximation distribution obtained by moment matching and the CDF of the sum of lognormal RVs.

In this work, we will derive a new bound on the CDF, and from this bound we will also derive new approximations to the CDF. The contributions of this work are summarized as:
\begin{itemize}
\item We utilize the lesser-known tangential mean-arithmetic mean inequality to derive a novel upper bound on the CDF for a sum of $N$ independent lognormal RVs. Our method replaces the sum of lognormal RVs with a product of shifted lognormal RVs. Remarkably, the bound is tight in the sense that it will converge to the desired CDF as the shift goes to infinity. 
\item We use the Mellin transform together with numerical integration in order to compute the bound. At the left tail of the CDF, we find that the bound can be expressed by a Q-function. 
\item We propose novel simple approximations to the CDF expressed by a product of Q-functions. The approximations are more accurate than existing ones, e.g., Farley's method.
\end{itemize}

Sec. II contains the derivation of the bound and a discussion about its tightness. Approximations to the CDF are contained in Sec. III and numerical evaluations are given in Sec. IV, while Sec. V concludes the paper.

\section{Bound on the CDF and its Left Tail Distribution}
\subsection{Preliminaries}
Let $X_i$ be a normal RV with probability density function (pdf)
\begin{equation}
f_{X_i}(x)=\frac{1}{\sqrt{2\pi}\sigma_i}e^{-\frac{(x-\mu_i)^2}{2\sigma_i^2}}, \,  -\infty<x<\infty, \, \sigma>0
\end{equation}
which implies that $e^{X_i}$ is a lognormal RV.  Our first objective is to give a bound for the CDF of the sum of $N$ lognormal RVs, 
\begin{equation}
S_N=\sum_{i=1}^N e^{X_i}. 
\end{equation}
The CDF is as ususal defined by $F_{S_N}(\gamma)=\int_0^\gamma f_{S_N}(x)dx$, where the pdf $f_{S_N}(x)$ does not exist on closed-form. In wireless communications, the outage probability is the probability that the SIR falls below a threshold, 
$\gamma_{\mathrm{th}}$. For lognormal shadow fading, where the received power is modelled by $e^{X_i}$, this can be expressed as
\begin{align}
\mathrm{Pr}\left [ \frac{e^{X_0}}{\sum_{i=1}^N e^{X_i}} \leq \gamma_{\mathrm{th}}\right]&=\mathrm{Pr}\left [ \sum_{i=1}^N e^{X_i} \geq \frac{e^{X_0}}{\gamma_{\mathrm{th}}} \right]\\
&=1-F_{S_N}\left(\frac{e^{X_0}}{\gamma_{\mathrm{th}}} \right ).
\end{align}
The l.h.s. is the complementary CDF (CCDF) of $S_N$.

The following results will be utilized for deriving the bound. For any set of positive integers, $\mathbf{y}=(y_1,y_2,\ldots,y_N)$, the arithmetic mean (AM), the geometric mean (GM) and the tangential mean\footnote{The general definition is $\mathrm{TM}(\delta)=(\delta+y_1)^{\alpha_1}\ldots(\delta+y_N)^{\alpha_N}-\delta$ where $\alpha_1+\alpha_2+\ldots+\alpha_N=1$.} (TM) are defined as:
\begin{align}
\mathrm{AM}(\mathbf{y})&=\frac{1}{N}\sum_{i=1}^N y_i\\
\mathrm{GM}(\mathbf{y})&=\left (\prod_{i=1}^Ny_i \right)^{\frac{1}{N}}\\
\mathrm{TM}(\mathbf{y},\delta)&=\left (\prod_{i=1}^N(\delta+y_i)\right)^{\frac{1}{N}}-\delta, \, \delta>0 \label{eq:TM}
\end{align}
It has been shown that the following inequality holds \cite{Sandor},
\begin{equation}
\mathrm{GM}(\mathbf{y})\le \mathrm{TM}(\mathbf{y},\delta)\le \mathrm{AM}(\mathbf{y}) \label{eq:amgm}
\end{equation}
where equality is achieved only when $y_1=y_2=\ldots=y_N$. Moreover, it was proven in \cite{Sandor} that TM$(\mathbf{y},\delta)$ is a monotonically increasing function of $\delta$, and it can be shown that (see Appendix)
\begin{equation}
\lim_{\delta\rightarrow \infty} \mathrm{TM}(\mathbf{y},\delta)=\mathrm{AM}(\mathbf{y}). \label{eq:tmas}
\end{equation}
Since $\mathrm{AM}=S_N/N$, the RV $N\cdot \mathrm{TM}(e^\mathbf{X},\delta)$ with $\mathbf{X}=(X_1,X_2,\ldots,X_N)$, will approach $S_N$ for large $\delta$.

\subsection{A Bound on the CDF}
Now, let us define the shifted lognormal RV, $Y_i=\delta+e^{X_i}$, which has a pdf given by 
\begin{equation}
f_{Y_i}(y,\delta)=\frac{1}{\sqrt{2\pi}\sigma_i(y-\delta)}e^{-\frac{(\ln(y-\delta)-\mu_i)^2}{2\sigma_i^2}}, \,  y>\delta, \delta\ge 0. \label{eq:shlg}
\end{equation}
The shift $\delta$ describes a linear translation of the standard lognormal pdf.
By definition $F_{Y_i}(\gamma,\delta)=\int_\delta^\gamma f_{Y_i}(y,\delta)dy$ and the CDF becomes 
\begin{equation}
F_{Y_i}(\gamma,\delta)=
\begin{cases}
0, & \gamma\le \delta \\
1-Q\left ( \frac{\ln(\gamma-\delta)-\mu_i}{\sigma_i}\right ), & \gamma>\delta
\end{cases}\label{eq:cdfY}
\end{equation}
where $Q(x)=1/(\sqrt{2\pi}\sigma)\int_x^\infty e^{-\frac{y^2}{2}}dy$. The moments can be defined from dB units by ${\sigma=\lambda\sigma_{\mathrm{dB}}}$ and ${\mu=\lambda\mu_{\mathrm{dB}}}$, with ${\lambda=\ln(10)/10\approx 0.23026}$ \cite{Beaulieu1}.
 
Moreover, define the product of shifted lognormal RVs, 
\begin{equation}
Z_N=\prod_{i=1}^NY_i, 
\end{equation}
and its CDF, 
\begin{equation}
F_{Z_N}(\gamma,\delta)=\int_{\delta^N}^\gamma f_{Z_N}(x,\delta)dx. 
\end{equation}
By using (\ref{eq:TM}) and ${\mathrm{AM}=S_N/N}$, our proposed upper bound on the CDF follows. 
\begin{equation}
F_{S_N}(\gamma) \le F_{Z_N}\left(\left(\frac{\gamma}{N}+\delta\right)^N,\delta\right), \, \gamma\geq 0 \label{eq:ub}
\end{equation}
It should be remarked that we have not used any independence assumption for the RVs to arrive at (\ref{eq:ub}). Hence, the bound holds also for correlated lognormal RVs. However, numerical evaluation of (\ref{eq:ub}) with correlated RVs appears to be a nontrivial task and we do not consider this case further herein.
\subsection{Tightness of the Bound}
Obviously due to (\ref{eq:amgm}),  (\ref{eq:ub}) is tighter than the bound in \cite{Berggren1}, which is based on the GM. Interestingly, (\ref{eq:ub}) is asymptotically tight, since (\ref{eq:tmas}) implies that 
\begin{equation}
\lim_{\delta\rightarrow \infty}F_{Z_N}\left(\left(\frac{\gamma}{N}+\delta\right)^N,\delta\right) =F_{S_N}(\gamma).
\end{equation}
The CDF can be obtained through (\ref{eq:cdf1}), which by variable substitution reduces to (\ref{eq:cdf2}). 
%\begin{figure*}[ht]
%\hrulefill
\begin{align}
\lim_{\delta\rightarrow \infty} F_{Z_N}\left(\left(\frac{\gamma}{N}+\delta\right)^N,\delta\right)&=\lim_{\delta\rightarrow \infty}\int_{\delta^N} ^{\left ( \frac{\gamma}{N}+\delta \right)^N} f_{Z_N}(x,\delta)dx\label{eq:cdf1}\\
&=\lim_{\delta\rightarrow \infty} \int_{0}^\gamma \left(\frac{x}{N}+\delta\right)^{N-1}
 f_{Z_N}\left(\left(\frac{x}{N}+\delta\right)^N,\delta\right) dx\label{eq:cdf2}
%F_{Z_2}\left(\left(\frac{\gamma}{2}+\delta\right)^2\right)&=\int_0^{\frac{\left(\frac{\gamma}{2}+\delta\right)^2}{\delta}-\delta}\left(1-
%Q\left(\frac{\ln\left(\frac{\left(\frac{\gamma}{2}+\delta\right )^2}{\delta+y}-\delta\right) -\mu}{\sigma}\right)\right)f_Y(y,0)dy \label{eq:cdfint}
\end{align}
%\hrulefill
%\end{figure*}
By taking the derivative w.r.t. to $\gamma$, it follows that the integrand in (\ref{eq:cdf2}) will converge to $f_{S_N}(x)$ as $\delta\rightarrow \infty$, i.e.,
\begin{equation}
f_{S_N}(x)=\lim_{\delta\rightarrow \infty} \left(\frac{x}{N}+\delta\right)^{N-1}
 f_{Z_N}\left(\left(\frac{x}{N}+\delta\right)^N,\delta\right).
\end{equation}

Furthermore, it is straightforward to show that for all the means
\begin{equation}
\min_i y_i \le \mathrm{GM}(\mathbf{y}), \mathrm{TM}(\mathbf{y},\delta),\mathrm{AM}(\mathbf{y})\le \max_i y_i. 
\end{equation}
Therefore, we have
\begin{equation}
\mathrm{AM}(\mathbf{y})-\mathrm{GM}(\mathbf{y})\le \max_i y_i-\min_i y_i \label{eq:bndlim}
\end{equation} 
which implies that the gap between the AM and the GM (and therefore also the gap between the AM and the TM) closes when the maximum difference between the elements in $\mathbf{y}$ approaches 0. Thus, firstly, the bound (\ref{eq:ub}) will become tighter with smaller variances $\sigma_i^2$, since the r.h.s of (\ref{eq:bndlim}) will then on average become smaller.  Secondly, for small $\gamma$, all $y_i$ will be small, i.e., $\epsilon=\max_i y_i-\min_i y_i$ will be small. Thus using (\ref{eq:amgm}), it follows from the squeeze theorem in calculus that  
\begin{align}
\lim_{\epsilon\rightarrow 0} \mathrm{AM}(\mathbf{y})-\mathrm{GM}(\mathbf{y})&=
\lim_{\epsilon\rightarrow 0} \mathrm{AM}(\mathbf{y})-  \mathrm{TM}(\mathbf{y},\delta)\label{eq:st}\\
&=0.
\end{align}
Hence, (\ref{eq:ub}) will approach the bound from \cite{Berggren1} for small $\gamma$, regardless of the $\delta$ value. 
If we define $V_i=e^{X_i}$ and $W_N=\prod_{i=1}^NV_i$, we have  the bound from \cite{Berggren1}
\begin{equation}
F_{W_N}(x)=1-Q\left(\frac{\ln x-\bar \mu}{\bar \sigma} \right)
\end{equation}
where $\bar \mu=\mathbb{E}[\sum_{i=1}^N X_i]=\sum_{i=1}^N \mu_i$ and $\bar \sigma^2=\mathbb{E}[\left(\sum_{i=1}^N X_i-\bar \mu\right)^2]$, where $\mathbb{E}$ is  the expectation value operator. Taking the limit, due to (\ref{eq:st}), we obtain
\begin{align}
\lim_{\gamma\rightarrow 0}F_{Z_N}\left(\left(\frac{\gamma}{N}+\delta\right)^N,\delta\right)&=\lim_{\gamma\rightarrow 0} F_{W_N}\left(\left(\frac{\gamma}{N}\right)^N\right )\\
&=\lim_{\gamma\rightarrow 0}1-Q\left( \frac{N\ln(\gamma/N)-\bar \mu}{\bar \sigma}\right). \label{eq:gmas}
\end{align}
We shall therefore expect that the bound (\ref{eq:ub}) is well approximated by  (\ref{eq:gmas}) for small $\gamma$, i.e., at the left tail of the CDF. 
We can then rewrite (\ref{eq:gmas}) for small $\gamma$ as
\begin{equation}
 F_{Z_N}(\gamma,\delta)\approx1-Q\left( \frac{\ln(\gamma)-\ln(N)-\bar \mu/N}{\bar \sigma/N}\right). \label{eq:gmas2}
\end{equation} 
and identify that the r.h.s. of (\ref{eq:gmas2}) describes the mean and variance of a lognormal RV with parameters  
\begin{align}
\hat \mu &=\ln(N)+\sum_{i=1}^N \mu_i/N \label{eq:log1}\\
\hat \sigma^2&=\sum_{i=1}^N \sigma^2_i/N^2. \label{eq:log2}
\end{align}

\subsection{Computation of the Bound}
The bound can be computed for the case of independent RVs as follows. The pdf for a product of independent RVs can be determined by multiplicative convolution (aka Mellin convolution) or by Mellin transforms \cite{Bertrand}. Since the $Y_i$ are independent, the pdf $f_{Z_N}(x)$ follows from $N$-times repeated Mellin convolution, and by defining $f_{Z_1}(y)=f_{Y}(y,\delta)$, and using the definition of Mellin convolution \cite{Bertrand}, we obtain: 
\begin{align}
f_{Z_n}(x)&=\int_{\delta^{n-1}}^{\frac{x}{\delta}}f_{Y}\left ( \frac{x}{y},\delta\right )\frac{f_{Z_{n-1}}(y)}{y} dy, x\ge \delta^n, \,n\ge 2 \label{eq:conv}
\end{align}
Thus, the CDF could be obtained by computing $N$ nested finite range integrals according to (\ref{eq:conv}). Alternatively, the computation could be done in the transform domain. The Mellin transform for (\ref{eq:shlg}) is defined for complex values $s=\alpha+\jj\beta$, where $\jj=\sqrt{-1}$, by
\begin{equation}
\phi_{Y}(s)=\int_\delta^\infty y^{s-1}f_{Y}(y,\delta)dy. \label{eq:mellin1}
\end{equation}
The largest interval $a<\alpha<b$ where (\ref{eq:mellin1}) converges is denoted as the fundamental strip. When $y\rightarrow \delta^+$ we have 
\begin{equation}
f_{Y}(y,\delta)\sim((y-\delta)(y-\delta)^{\ln(y-\delta)})^{-1}
\end{equation}
and when $y\rightarrow \infty$, we have 
\begin{equation}
f_{Y}(y,\delta)\sim(yy^{\ln(y)})^{-1} 
\end{equation}
so 
\begin{equation}
f_{Y}(y,\delta)\underset{y\rightarrow \delta^+}{=} \mathcal{O}((y-\delta)^{-\ln(y-\delta)}), \, f_{Y}(y,\delta)\underset{y\rightarrow \infty}{=} \mathcal{O}(y^{-\ln(y)})
\end{equation}
and since $\lim_{y\rightarrow \delta^+}\ln(y-\delta)=-\infty$  and $\lim_{y\rightarrow\infty}\ln(y)=\infty$, we have $a=-\infty$ and $b=\infty$ (cf. Lemma 1 \cite{Flajolet}). The Mellin transform of the $N$-times Mellin convolution (\ref{eq:conv}) is the product of the $N$ Mellin transforms \cite{Bertrand}. The inversion of the Mellin transform is unique and is given by the line integral for any $\alpha$ where $a<\alpha<b$
\begin{equation}
f_{Z_N}(x)=\frac{1}{2\pi \jj}\int_{\alpha-\jj\cdot\infty}^{\alpha+\jj\cdot\infty}x^{-s}\phi_{Y}(s)^N ds \label{eq:mellin2}
\end{equation}
which is thus admitted for any real value $\alpha$. There is no closed-form expression for $\phi_{Y}(s)$ and numerical integration of (\ref{eq:mellin1}) and (\ref{eq:mellin2}) will be required to determine $F_{Z_N}\left(\left(\frac{\gamma}{N}+\delta\right)^N\right)$. For convenience of notation, (\ref{eq:conv}) and (\ref{eq:mellin2}) are given for $\mu_i=\mu$ and $\sigma_i=\sigma$ but it is straightforward to generalize to non-uniform parameters $\mu_i$ and $\sigma_i$.

\begin{figure}
\begin{center}
\includegraphics[width=\textwidth]{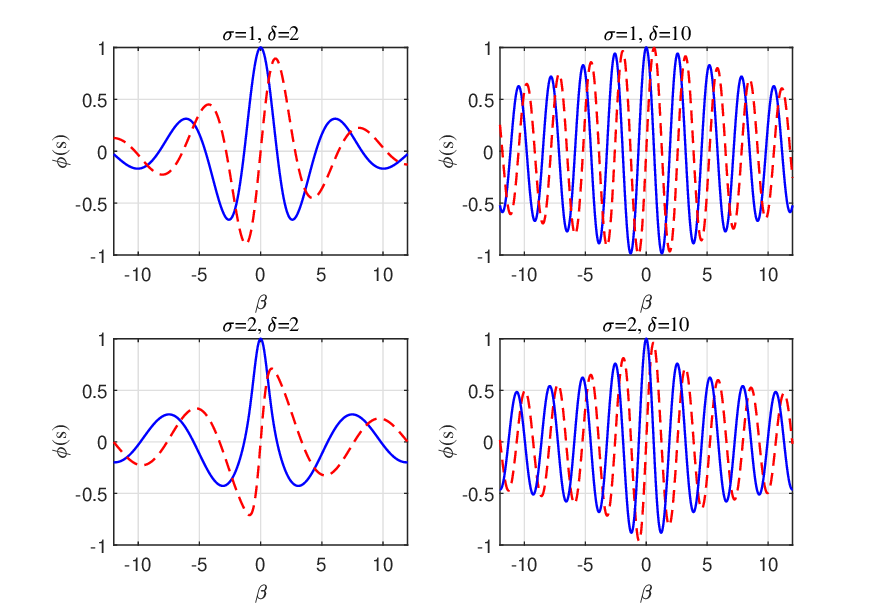}
\caption{The real part (solid blue) and imaginary part (dashed red) of the Mellin transform $\phi(\alpha+\jj\beta)$  for $\alpha=1$, for $\delta=2$ and $\delta=10$, $\sigma=1$ and $\sigma=2$, with $\mu=0$.}
\label{fig:4plots}
\end{center}
\vspace{-7 mm}
\end{figure}

Fig. \ref{fig:4plots} depicts the Mellin transform\footnote{In our evaluations, numerical integration by an adaptive quadrature method is used for (\ref{eq:mellin1}) and a trapezoidal method is used for (\ref{eq:mellin2}).} as function of $\beta$ for a fixed value of $\alpha$. The decay of $\phi(s)$ becomes slower as $\delta$ increases, which implies that a larger range of $\beta$ values needs to be evaluated in order to have sufficiently small integration error in (\ref{eq:mellin2}).  

%\begin{figure*}[ht]
%\hrulefill
%\begin{align}
%F_{Z_2}\left(\left(\frac{\gamma}{2}+\delta\right)^2\right)&=\int_\delta^{\frac{\left( \frac{\gamma}{2}+\delta\right )^2}{\delta}}F_{Y}\left( \frac{1}{y}\left(\frac{\gamma}{2}+\delta \right)^2,\delta \right)f_{Y}(y,\delta)dy\\\
%&=\int_0^{\frac{\left(\frac{\gamma}{2}+\delta\right)^2}{\delta}-\delta}\left(1-
%Q\left(\frac{\ln\left(\frac{\left(\frac{\gamma}{2}+\delta\right )^2}{\delta+y}-\delta\right) -\mu}{\sigma}\right)\right)f_Y(y,0)dy \label{eq:cdfint}\\
%\lim_{\delta\rightarrow \infty} F_{Z_2}\left(\left(\frac{\gamma}{2}+\delta\right)^2\right)&=\int_{-\infty}^{\ln(\gamma)}\underbrace{ \left(1- Q\left(\frac{\ln(\gamma-e^{x})-\mu}{\sigma} \right) \right)}_{g(x)}f_X(x)dx \label{eq:cdfint2}
%\end{align}
%\hrulefill
%\end{figure*}

\section{Approximations of the CDF}
We next seek to find simple approximations to the r.h.s. of (\ref{eq:ub}). These expressions cannot be formally proven to be upper bounds on the CDF, although they may practically become so for a large range of $\gamma$ values. Hence, they are referred to as approximations of the CDF.
\subsection{Approximation for $N=2$} 
For $N=2$, we have $Z_2=Y_1Y_2$ and the CDF becomes
\begin{equation}
F_{Z_2}(x)=\int_\delta^{\frac{x}{\delta}}F_Y\left ( \frac{x}{y},\delta\right ) f_Y(y,\delta)dy, \,x\ge\delta^2
\end{equation}
where we for convenience of notation have assumed $\sigma_i=\sigma$ and $\mu_i=\mu, \,\forall i$. 
Therefore, we obtain
\begin{align}
F_{Z_2}\left(\left(\frac{\gamma}{2}+\delta\right)^2\right)&=\int_\delta^{\frac{\left( \frac{\gamma}{2}+\delta\right )^2}{\delta}}F_{Y}\left( \frac{1}{y}\left(\frac{\gamma}{2}+\delta \right)^2,\delta \right)f_{Y}(y,\delta)dy\\
&=\int_0^{\frac{\left(\frac{\gamma}{2}+\delta\right)^2}{\delta}-\delta}\left(1-
Q\left(\frac{\ln\left(\frac{\left(\frac{\gamma}{2}+\delta\right )^2}{\delta+y}-\delta\right) -\mu}{\sigma}\right)\right)f_Y(y,0)dy. \label{eq:cdfint}
\end{align}
where (\ref{eq:cdfY}) was used in (\ref{eq:cdfint}). Since 
\begin{equation}
\lim_{\delta\rightarrow \infty} \frac{\left(\frac{\gamma}{2}+\delta \right)^2}{y+\delta}-\delta=\gamma-y
\end{equation}
it is straightforward to verify that (\ref{eq:cdfint}) asymptotically reduces to $F_{S_2}(\gamma)=\int_0^\gamma F_{Y}(\gamma-y,0)f_{Y}(y,0)dy$, i.e., the CDF resulting from the integration for the CDF obtained by a convolution of $f_{Y}(y,0)$. By change of integration variable, we have the equivalent representation 
\begin{equation}
\lim_{\delta\rightarrow \infty} F_{Z_2}\left(\left(\frac{\gamma}{2}+\delta\right)^2\right)=\int_{-\infty}^{\ln(\gamma)}\underbrace{ \left(1- Q\left(\frac{\ln(\gamma-e^{x})-\mu}{\sigma} \right) \right)}_{g(x)}f_X(x)dx. \label{eq:cdfint2}
\end{equation}
Focusing on the case $\delta \rightarrow \infty$, where the bound is tight, an approximation to the CDF can be obtained on closed-form (i.e., not requiring integration) from (\ref{eq:cdfint2})  by the conventional approximation $\mathbb{E}g(x)\approx g(\mathbb{E}x)$, where the expectation operator $\mathbb{E}$ is with respect to the RV $x$ with pdf $h_2(x)=f_X(x)/C_2(\gamma)$. The normalization factor
\begin{align}
C_2(\gamma)&=\int_{-\infty}^{\ln (\gamma)}f_X(x)dx\\
&=1-Q\left( \frac{\ln(\gamma)-\mu}{\sigma}\right)
\end{align}
is needed such that $h_2(x)$ is a pdf on the interval $-\infty<x\le \ln(\gamma)$. 
Then, we get the approximation
\begin{align}
F_{S_2}(\gamma)\approx C_2(\gamma) \left ( 1-Q\left( \frac{\ln(\gamma-e^{\mu_2(\gamma)})-\mu}{\sigma}\right) \right ) \label{eq:bnd}
\end{align}
where $\mu_2=\mathbb{E}x$, which equals
\begin{align}
\mu_2(\gamma)&=\int_{-\infty}^{\ln(\gamma)} xh_2(x)dx\\
&= \frac{1}{C_2(\gamma)}\left( \mu\left( 1-Q\left( \frac{\ln(\gamma)-\mu}{\sigma}\right ) \right) -\frac{\sigma}{\sqrt{2\pi}}e^{-\frac{(\ln(\gamma)-\mu)^2}{2\sigma^2}}\right)\\
&=\mu-\frac{\sigma}{\sqrt{2\pi}C_2(\gamma)}e^{-\frac{(\ln(\gamma)-\mu)^2}{2\sigma^2}}
\end{align}
An interpretation of (\ref{eq:bnd}) is that the CDF is the product between the CDF of the lognormal RV and the CDF of a shifted lognormal RV, where the shift $e^{\mu_2(\gamma)}$ is not fixed but depends on $\gamma$.

To assess whether the r.h.s. of (\ref{eq:bnd}) may become an upper bound, we inspect the derivatives of $g(x)$,
\begin{align}
\frac{dg(x)}{dx}&=-\frac{e^{x-\frac{(\ln(\gamma-e^x)-\mu)^2}{2\sigma^2}}}{\sqrt{2\pi}\sigma(\gamma-e^x)}\\
\frac{d^2g(x)}{dx^2}&=-\frac{e^{x-\frac{(\ln(\gamma-e^x)-\mu)^2}{2\sigma^2}}(\gamma\sigma^2+e^x\ln(\gamma-e^x)-\mu e^x))}{\sqrt{2\pi}\sigma^3(\gamma-e^x)^2}
\end{align}
leading to that $dg(x)/dx$ is negative on the interval $-\infty<x\le \ln(\gamma)$ and that it is a monotonically non-increasing function for  $-\infty<x\le x_0$, where $x_0$ is the solution to 
\begin{equation}
\gamma\sigma^2+e^{x_0}\ln(\gamma-e^{x_0})-\mu e^{x_0}=0.
\end{equation}
Thus, $g(x)$ is a concave function on the interval $-\infty<x\le x_0$.
Therefore, if $\ln(\gamma)-x_0$ is relatively small, the r.h.s. of (\ref{eq:bnd}) may practically become an upper bound for a large range of $\gamma$ values since Jensen's inequality implies that $\mathbb{E}g(x)\le g(\mathbb{E}x)$ for $-\infty<x\le x_0$, albeit $g(x)$ is not concave on $x_0<x\le \ln(\gamma)$.

\subsection{Approximation for $N>2$}
The approximation method of Sec. III.A can be applied repeatedly for $N>2$ and we will show this explicitly for $N=3$. By defining $\gamma_2=\gamma-e^{\mu_2(\gamma)}$ and using $S_3=S_2+e^{X_3}$ with (\ref{eq:bnd}), it follows that 
\begin{align}
F_{S_3}(\gamma)&\approx C_2(\gamma)\!\!\!\!\!\!\int\displaylimits_{-\infty}^{\ln(\gamma_2)}\left(1-Q\left(\frac{\ln(\gamma_2-e^x)-\mu}{\sigma} \right) \right )f_X(x)dx \label{eq:app2a}\\
&\approx C_3(\gamma)C_2(\gamma)\left(1-Q\left(\frac{\ln(\gamma_2-e^{\mu_3(\gamma)})-\mu}{\sigma} \right) \right) \label{eq:app2}
\end{align}
where the approximation in (\ref{eq:app2a}) is due to the approximation of $F_{S_2}(\gamma)$ and (\ref{eq:app2}) is due to $\mathbb{E}g(x)\approx g(\mathbb{E}x)$, where the RV $x$ has a pdf $h_3(x)=f_X(x)/C_3(\gamma)$, and the normalization factor
\begin{align}
C_3(\gamma)&=\int\displaylimits_{-\infty}^{\ln (\gamma_2)}f_X(x)dx\\
&=1-Q\left( \frac{\ln(\gamma_2)-\mu}{\sigma}\right)
\end{align}
is needed such that $h_3(x)$ is a pdf on the interval $-\infty<x\le \ln(\gamma_2)$. Furthermore, we obtain
\begin{align}
\mu_3(\gamma)&=\int\displaylimits_{-\infty}^{\ln(\gamma_2)} xh_3(x)dx\\
&= \frac{1}{C_3(\gamma)}\left( \mu\left( 1-Q\left( \frac{\ln(\gamma_2)-\mu}{\sigma}\right ) \right) -\frac{\sigma}{\sqrt{2\pi}}e^{-\frac{(\ln(\gamma_2)-\mu)^2}{2\sigma^2}}\right)\\
&=\mu-\frac{\sigma}{\sqrt{2\pi}C_3(\gamma)}e^{-\frac{(\ln(\gamma_2)-\mu)^2}{2\sigma^2}}
\end{align}
The error in the approximation is due to reusing the pdf for $N=2$ in (\ref{eq:app2a}) and thereafter using $\mathbb{E}g(x)\approx g(\mathbb{E}x)$ again. Thus, the approximation may be less accurate as $N$ increases. The above steps can be applied repeatedly to obtain $\gamma_N$, $\mu_N$ and $C_N$ for $N>3$, which could be used to derive an approximation as for (\ref{eq:app2}).

\subsection{Approximation for large $N$}
Let us define the RV $\tilde Z_N=Z_N^{\frac{1}{N}}$ and assume that $N$ is large. Since
\begin{align}
\ln \left(\tilde Z_N\right)= \frac{1}{N}\sum_{i=1}^N \ln (Y_i),
\end{align}
a central limit theorem (CLT) implies that, as $N\rightarrow \infty$, $\ln (\tilde Z_N)$ will be a normal RV with mean $\tilde\mu$ and variance $\tilde\sigma^2$. Therefore, $\tilde Z_N$ will converge to a lognormal RV. 
The mean and variance can be determined according to the CLT through
\begin{align}
%\tilde\mu&=\int_\delta^\infty \ln(y)f_Y(y,\delta)dy\\
\tilde\mu&=\int_{-\infty}^\infty \ln(\delta+e^x)f_X(x)dx\\
&\approx \frac{1}{\sqrt{\pi}}\sum_{m=1}^Mw_m \ln(\delta+e^{\sqrt{2}\sigma x_m+\mu}) \label{eq:gh1}\\
\tilde\sigma^2&=\frac{1}{N}\left (\int_{-\infty}^\infty \ln(\delta+e^x)^2f_X(x)dx-\tilde\mu^2 \right )\\
&\approx\frac{1}{N}\left (\frac{1}{\sqrt{\pi}}\sum_{m=1}^Mw_m \ln(\delta+e^{\sqrt{2}\sigma x_m+\mu})^2 -\tilde\mu^2 \right )\label{eq:gh2}
\end{align}
where (\ref{eq:gh1}) and (\ref{eq:gh2}) are due to Gauss-Hermite integration of order $M$. The weights $w_m$ and abscissas $x_m$ can be found in \cite{Stegun}. By using (\ref{eq:TM}) and AM=$S_N/N$, we can therefore make the following approximation for large $N$.
\begin{align}
F_{Z_N}\left(\left(\frac{\gamma}{N}+\delta\right)^N\right)&\approx F_{\tilde Z_N}\left(\frac{\gamma}{N}+\delta\right)\\
&=1-Q\left ( \frac{\ln\left(\frac{\gamma}{N}+\delta\right)-\tilde\mu}{\tilde\sigma}\right ) \label{eq:appro}
\end{align}

\section{Numerical Examples}
\subsection{Evaluation of the Left Tail}
\begin{figure}
\begin{center}
\includegraphics[width=\textwidth]{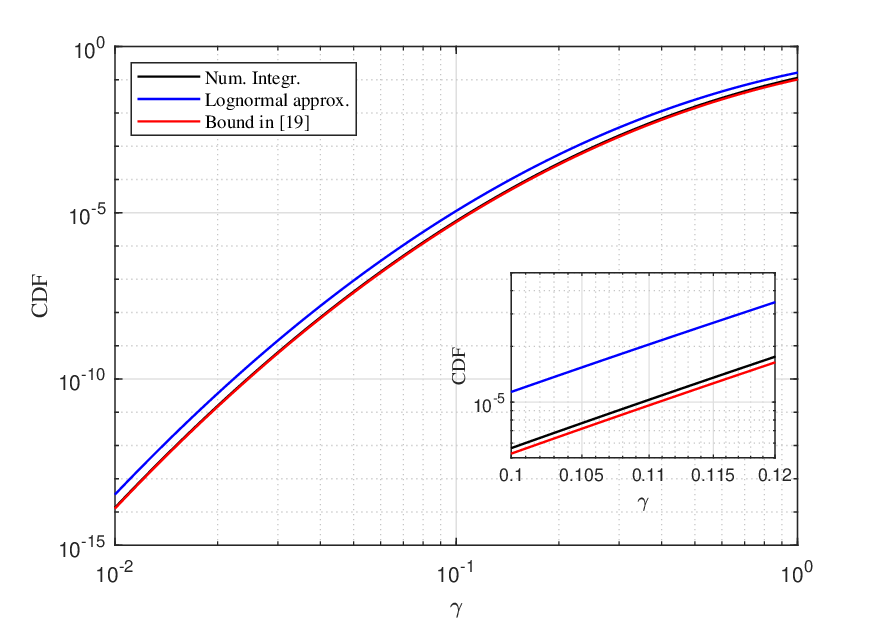}
\caption{The CDF obtained by numerical integration and the lognormal approximation for $N=2$, $\sigma=1$ and $\mu=0$.}
\label{fig:as}
\end{center}
\vspace{-7 mm}
\end{figure}

Fig. \ref{fig:as} shows the l.h.s. of (\ref{eq:gmas2}) and the CDF, which is obtained by numerical integration, for the case of $N=2$, $\sigma=1$ and $\mu=0$. The distance between the curves is almost constant, but since the plot is in logarithmic scale, it implies that the actual difference between the two curves decreases as $\gamma$ decreases. Hence, it confirms the bound (\ref{eq:ub}) can for small $\gamma$ be approximated by the  bound of \cite{Berggren1}. 
However, asymptotically as $\gamma\rightarrow 0$, the distribution of $S_N$ may not be that of a lognormal RV \cite{Zhu2}. As a reference case the lower bound [(10a)-(10c) of \cite{Beaulieu2}], which applies for $N=2$ and is based on the Marcum-Q function is included, which is even tighter.  
\subsection{Evaluation of the Bound}
\begin{figure}
\begin{center}
\includegraphics[width=\textwidth]{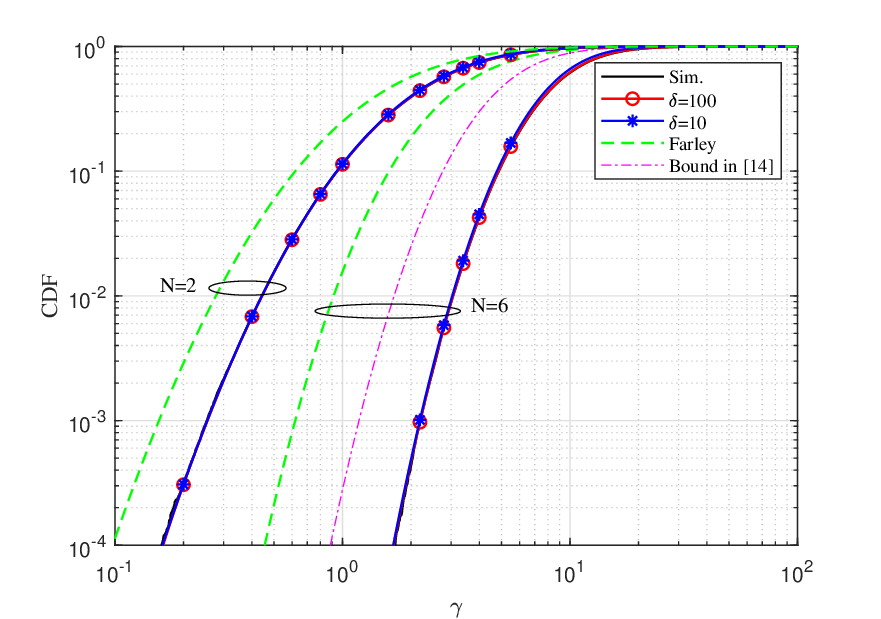}
\caption{CDF for the proposed bound (\ref{eq:ub}) with $\delta=10$ and $\delta=100$, $N=2$ and $N=6$, for $\sigma=1$ with $\mu=0$.}
\label{fig:cdf}
\end{center}
\vspace{-7 mm}
\end{figure}

We plot the CCDF, hence the hitherto derived upper bounds on the CDF are displayed as lower bounds on the CCDF. Additionally, the CDF is plotted in Fig. \ref{fig:cdf} in order to be able to compare bounds at low $\gamma$. For comparison, we use previously published lower bounds on the CCDF. For example, the r.h.s. of the following inequality is plotted
\begin{equation}
 1-F_{S_N}(\gamma) \ge 1-\left ( 1 -Q\left ( \frac{\ln(\gamma)-\mu}{\sigma}\right )\right)^N\label{eq:farley}
\end{equation}
which is a special case of \cite{Slimane} and is also referred to as Farley's method \cite{Beaulieu0}. Notably, the bound of \cite{Berggren1} is based on the GM and it will be worse than (\ref{eq:ub}), so it is not included. For $N=6$, we also include the improved bound [(eq. (11) in \cite{Slimane}], which is a two-dimensional integral that is evaluated numerically. 
The distribution of $S_N$ obtained by Monte-Carlo simulations is thereto contained in the plots. 

\begin{figure}
\begin{center}
\includegraphics[width=\textwidth]{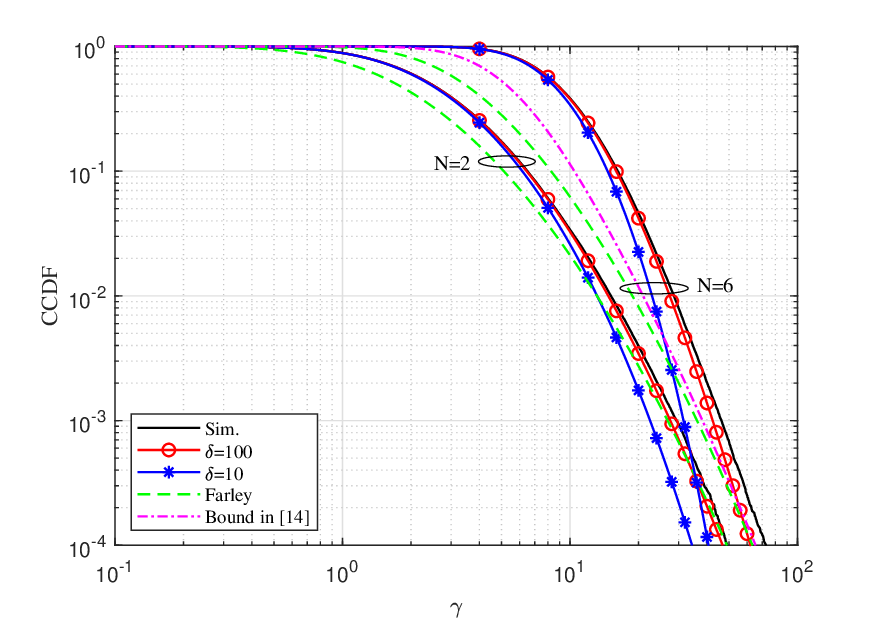}
\caption{CCDF for the proposed bound (\ref{eq:ub}) with $\delta=10$ and $\delta=100$, $N=2$ and $N=6$, for $\sigma=1$ with $\mu=0$.}
\label{fig:ccdf1}
\end{center}
\vspace{-7 mm}
\end{figure}

\begin{figure}
\begin{center}
\includegraphics[width=\textwidth]{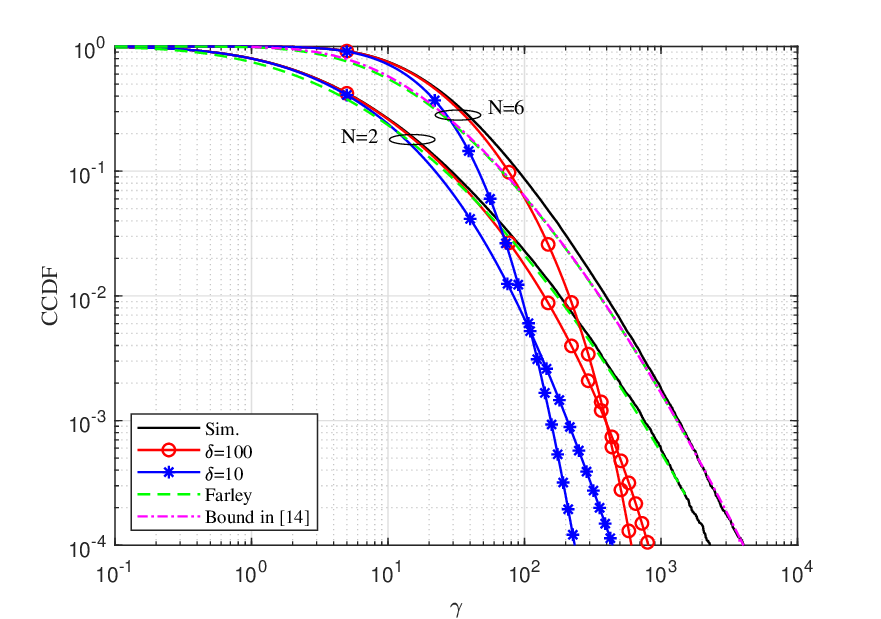}
\caption{CCDF for the proposed bound (\ref{eq:ub}) with $\delta=10$ and $\delta=100$, $N=2$ and $N=6$, for $\sigma=2$ with $\mu=0$.}
\label{fig:ccdf2}
\end{center}
\vspace{-7 mm}
\end{figure}

We select $\delta=10$ and $\delta=100$ in order to demonstrate the bounds, and apply $\sigma=1\,(\sigma_{\mathrm{dB}}=4.34)$ and $\sigma=2\,(\sigma_{\mathrm{dB}}=8.69)$. Typical vaules of $\sigma_{\mathrm{dB}}$ in radio channels are in the range 4--12 dB. From Fig. \ref{fig:ccdf1} and Fig. \ref{fig:ccdf2}, as expected, it can be observed that the bound is tighter for smaller $\sigma$. Morover, as $N$ increases, a larger value of $\delta$ is required in order to close the gap to the desired CCDF, especially for large $\sigma$. For small $\gamma$, the bound is accurate in general.  For the evaluated values of $\delta$, the new bounds are shown to be better than (\ref{eq:farley}) for small $\gamma$, while (\ref{eq:farley}) is better for large $\gamma$. Fig. \ref{fig:cdf} shows that the proposed bound is very accurate at the left tail of the CDF, whereas the previous bounds are not good at all.

\subsection{Evaluation of the Approximations}
\begin{figure}
\begin{center}
\includegraphics[width=\textwidth]{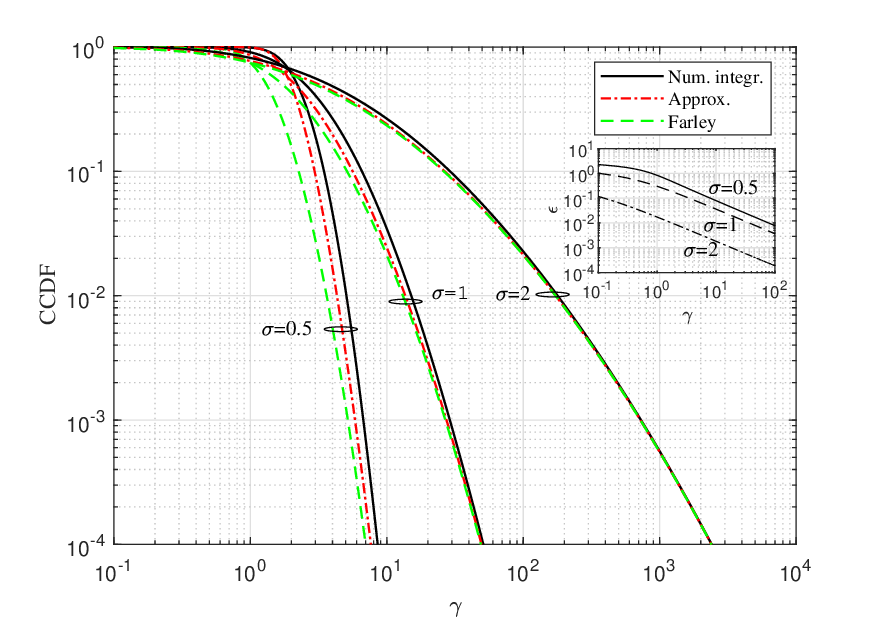}
\caption{CCDF of approximation (\ref{eq:bnd}) for $N=2$ for $\sigma=0.5, 1$ and 2 with $\mu=0$. The inset shows the difference $\epsilon=\ln(\gamma)-x_0$ for different $\sigma$.}
\label{fig:ccdf3}
\end{center}
\vspace{-7 mm}
\end{figure}

\begin{figure}
\begin{center}
\includegraphics[width=\textwidth]{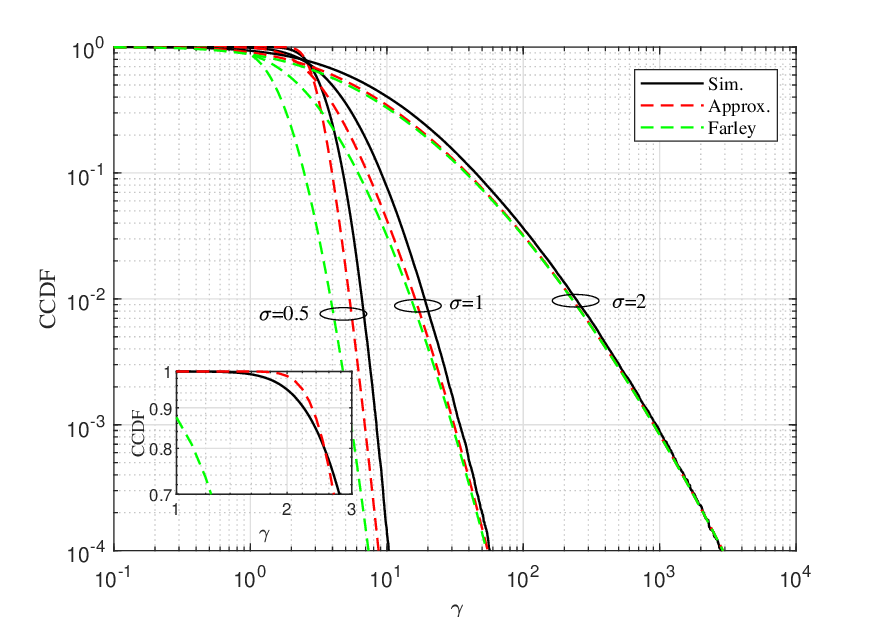}
\caption{CCDF of approximation (\ref{eq:bnd}) for $N=3$ for $\sigma=0.5, 1$ and 2 with $\mu=0$. The inset shows a zoom in for the case of $\sigma=0.5$.}
\label{fig:ccdf3b}
\end{center}
\vspace{-7 mm}
\end{figure}

Fig. \ref{fig:ccdf3} contains the approximation for $N=2$ and shows that (\ref{eq:bnd}) appears as a lower bound and that it is better than Farley's method (\ref{eq:farley}). In particular, the difference is significant for small $\sigma$. We solve for $x_0$ numerically and the subplot includes $\epsilon= \ln(\gamma)-x_0$, which shows that $\epsilon$ decreases with increasing $\gamma$ and increasing $\sigma$. If $\epsilon$ is fairly small, $g(x)$ is concave over most of the integration interval which leads to that (\ref{eq:bnd}) practically appears as an upper bound. Fig. \ref{fig:ccdf3b} shows that the approximation (\ref{eq:app2}) for $N=3$ is accurate for large $\gamma$, i.e., the right tail, and for large $\sigma$. The inset plot shows that for small $\gamma$ and small $\sigma$, the approximation (\ref{eq:app2}) intersects with the CDF and it is thus not a lower bound. These results show that the suggested approximation works well, at least when $N$ is moderately large.

\begin{figure}
\begin{center}
\includegraphics[width=\textwidth]{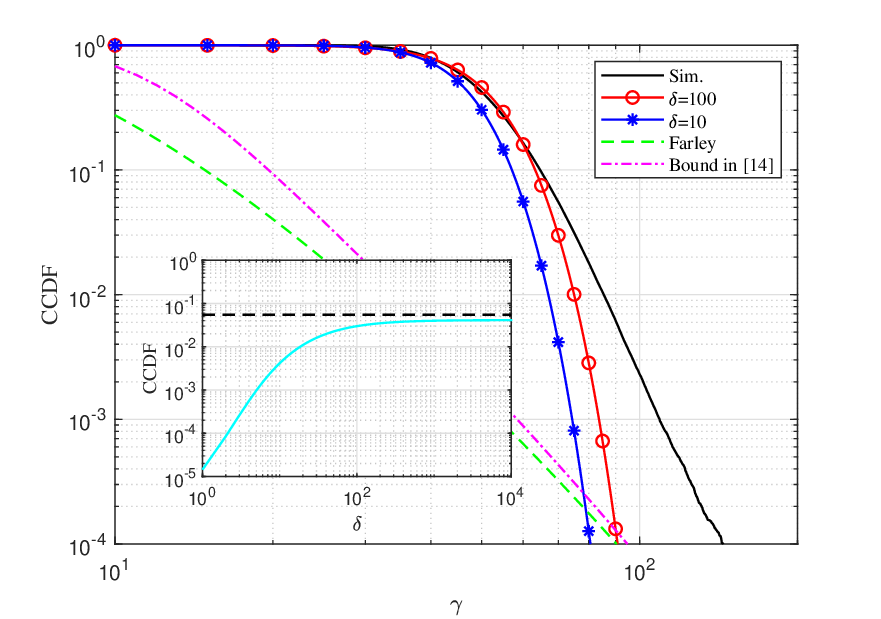}
\caption{CCDF of approximation (\ref{eq:appro}) for $\delta=10$ and $\delta=100$, for $N=30$, $\sigma=1$ and $\mu=0$. The inset shows the CCDF for $\gamma=70$ as function of $\delta$ and where the dashed line is the value obtained by simulation.}
\label{fig:ccdf4}
\end{center}
\vspace{-7 mm}
\end{figure}

For the case of large $N$, since the product of $N$ RVs is replaced with a single RV, the inequality (\ref{eq:amgm}) cannot be used and the approximation (\ref{eq:appro}) is not necessarily a lower bound on the CCDF for an arbitrary $\delta$. This can be observed from Fig. \ref{fig:ccdf4}, wherein the curve for $\delta=100$ intersects with the desired CCDF, where we have used $N=30$. The inset plot shows that, for the fixed value $\gamma=70$, (\ref{eq:appro}) is an increasing function of $\delta$ but it does not converge to the actual value of the CCDF. Notably, the previous bounds are not accurate at all for such large $N$ and do not work well.

\section{Conclusions}
We have derived an upper bound on the CDF of a sum of lognormal RVs which becomes tight and converges to the CDF for large values of the shift $\delta$. Thus, it is more accurate as $\delta$ increases and will outperform previously suggested upper bounds and approximation methods. Evaluation is done by numerical integration. The price of its accuracy is that it may require a larger computational cost in terms of numerical integration effort, than approximation methods, e.g., the ones based on moment matching or those using other pdfs than the lognormal, or other bounds. The bound can be approximated by a single Q-function at the left tail. Furthermore, we gave simple approximations to the bound on closed-form by means of products of Q-functions.  These approximations were shown to be more accurate than, e.g., Farley's method, at least for moderately large $N$. An implication of this work is that the classical problem of the sum of lognormal RVs could alternatively be viewed as a problem of a product of shifted lognormal RVs with large shifts.

%\newpage

\appendix
First, rewrite the $N$th root:
\begin{align}
\left (\prod_{i=1}^N(\delta+y_i)\right)^{\frac{1}{N}}&=\delta \left (1+\frac{y_1}{\delta}\right)^{\frac{1}{N}}
\left (1+\frac{y_2}{\delta}\right)^{\frac{1}{N}}\ldots\left (1+\frac{y_N}{\delta}\right)^{\frac{1}{N}}\label{eq:a1}
\end{align}
Furthermore, the Taylor series at $t=0$ can be obtained as
\begin{equation}
\left (1+t\right)^{\frac{1}{N}}=1+\frac{t}{N}+\frac{1}{N}\left(\frac{1}{N}-1\right)\frac{t^2}{2}+\mathcal{O}(t^3).\label{eq:taylor}
\end{equation}
Utilizing the first two terms of (\ref{eq:taylor}) and expanding (\ref{eq:a1}), we obtain
\begin{align}
\lim_{\delta\rightarrow \infty}\left (\prod_{i=1}^N(\delta+y_i)\right)^{\frac{1}{N}}\!\!\!\!-\delta =&
\lim_{\delta\rightarrow \infty}\delta\left(1+\frac{y_1}{N\delta}+\mathcal{O}\left (\frac{1}{\delta^2}\right) \right)
\left(1+\frac{y_2}{N\delta}+\mathcal{O}\left (\frac{1}{\delta^2} \right) \right)\ldots
\left(1+\frac{y_N}{N\delta}+\mathcal{O}\left (\frac{1}{\delta^2}\right) \right)-\delta\\
=&\lim_{\delta\rightarrow \infty}\frac{y_1+y_2+\ldots+y_N}{N}+\mathcal{O}\left (\frac{1}{\delta^2}\right )\\
=&\frac{y_1+y_2+\ldots+y_N}{N}.
\end{align}


\begin{thebibliography}{1}
\bibitem{Beaulieu0} N. C. Beaulieu, A. A. Abu-Dayya, and P. J. McLane, ``Estimating the distribution of a sum of independent lognormal random variables,"  \emph{ IEEE Trans. Commun.}, vol. 43, no. 12, pp. 2869--2873, Dec. 1995. 
\bibitem{Fenton} L. F. Fenton, ``The sum of lognormal probability distributions in scatter transmission systems,'' \emph{IRE Trans. Commun. Syst.}, vol. 8, no. , pp. 57--67, 1960.
\bibitem{Schwarz} S. Schwartz and Y. Yeh, ``On the distribution function and moments of power sums with lognormal components,'' \emph{Bell. Syst. Tech. J.}, vol. 61, pp. 1441--1462, 1982.
\bibitem{Tellambura1} C. Tellambura and D. Senaratne, ``Accurate computation of the MGF of the lognormal distribution and its application to sum of lognormals,'' \emph{IEEE Trans. Commun.}, vol. 58, no. 5, pp. 1568--1577, May 2010.
\bibitem{Mehta} N. B. Mehta, J. Wu, A. F. Molisch, and J. Zhang, ``Approximating a sum of lognormal random variables with a lognormal,'' \emph{IEEE Trans. Wireless Commun.}, vol. 6, no. 7, pp. 2690--2699, July 2007.
\bibitem{Beaulieu1} N. C. Beaulieu and Q. Xie, ``An optimal lognormal approximation to lognormal sum distributions,'' \emph{IEEE Trans. Veh. Technol.}, vol. 53, no. 2, pp. 479--489, Mar. 2004.
\bibitem{Miles} J. Miles, ``On the Laplace transform of the lognormal distribution: analytic continuation and series approximations,'' \emph{J. of Computational and Applied Math.}, vol. 404, Apr. 2022.
\bibitem{Zhao} L. Zhao and J. Ding, ``Least squares approximations to lognormal sum distributions," \emph{IEEE Trans. Veh. Technol.}, vol. 56, no. 2, pp. 991--997, Mar. 2007.
\bibitem{Nie} H. Nie and S. Chen, ``Lognormal sum approximation with type IV Pearson distribution," \emph{IEEE Commun. Lett.}, vol. 11, no. 10, pp. 790--792, Oct. 2007.
\bibitem{Liu} Z. Liu, J. Almhana, and R. McGorman, ``Approximating lognormal sum distributions with power lognormal distributions," \emph{IEEE Trans. Veh. Technol.}, vol. 57, no. 4, pp. 2611--2617, July 2008.
\bibitem{DiRenzo} M. Di Renzo, F. Graziosi, and F. Santucci, ``Further results on the approximation of log-normal power sum via Pearson type IV distribution: a general formula for log-moments computation,'' \emph{IEEE Trans. Commun.}, vol. 57, no. 4, pp. 893--898, Apr. 2009.
\bibitem{Li} X. Li, Z. Wu, V. D. Chakravarthy, and Z. Wu, ``A low-complexity approximation to lognormal sum distributions via transformed log skew normal distribution," \emph{IEEE Trans. Veh. Technol.}, vol. 60, no. 8, pp. 4040--4045, Oct. 2011.
\bibitem{Lam} C. L. J. Lam and T. Le-Ngoc, ``Log-shifted gamma approximation to lognormal sum distributions," \emph{IEEE Trans. Veh. Technol.}, vol. 56, no. 4, pp. 2121--2129, July 2007.
\bibitem{Slimane} S. B. Slimane, ``Bounds on the distribution of a sum of independent lognormal random variables,'' \emph{IEEE Trans. Commun.}, vol. 49, no. 6, pp. 975--978, June 2001.
\bibitem{Tellambura} C. Tellambura, ``Bounds on the distribution of a sum of correlated lognormal random variables and their application," \emph{IEEE Trans. Commun.}, vol. 56, no. 8, pp. 1241--1248, Aug. 2008.
\bibitem{Berggren1} F. Berggren and S. B. Slimane, ``A simple bound on the outage probability with lognormally distributed interferers,'' \emph{IEEE Commun. Lett.}, vol. 8, no. 5, pp. 271--273, May 2004.
\bibitem{Zhu} B. Zhu, Z. Zhang, L. Wang, J. Dang, L. Wu, J. Cheng, and G. Li, ``Right tail approximation for the distribution of lognormal sum and its applications,'' in \emph{Proc. IEEE Globecom}, Taipei, Taiwan, 2020, pp. 1--6.
\bibitem{Xiao} Z. Xiao, B. Zhu, J. Cheng, and Y. Wang, ``Outage probability bounds of EGC over dual-branch non-identically distributed independent lognormal fading channels with optimized parameters,'' \emph{IEEE Trans. Veh. Technol.}, vol. 68, no. 8, pp. 8232--8237, Aug. 2019. 
\bibitem{Beaulieu2} N. C. Beaulieu and G. Luan, ``On the Marcum Q-function behavior of the left tail probability of the lognormal sum distribution," in \emph{ Proc. IEEE ICC}, Dublin, Ireland, 2020, pp. 1--6.
\bibitem{Berggren2} F. Berggren ``An error bound for moment matching methods of lognormal sum distributions,'' \emph{European Trans. Telecommun.}, vol. 16, no. 6, pp. 573--577, 2005.
\bibitem{Sandor} J. S\'{a}ndor, \emph{Theory of means and their inequalities} [Online]. Available: https://www.math.ubbcluj.ro/$\sim$jsandor/ 
%\bibitem{Szyszkowicz} S. S. Szyszkowicz and H. Yanikomeroglu, ``On the tails of the distribution of the sum of lognormals,” in \emph{Proc. IEEE ICC}, Glasgow, UK, 2007, pp. 5324-5329.
\bibitem{Bertrand} J. Bertrand, P. Bertrand, and J.-P. Ovarlez, ``Chapter 12: The Mellin transform,''
\emph{The transform and applications handbook}, Ed. A.D. Poularikas, CRC Press inc, 1995.
\bibitem{Flajolet} P. Flajolet, X. Gourdon, and P. Dumas, ``Mellin transforms and asymptotics: harmonic sums,'' \emph{Theoretical Computer Science},
vol. 144, no. 1--2, pp. 3--58, June 1995.
\bibitem{Stegun} M. Abramowitz and J. Stegun, \emph{Handbook of mathematical functions with formulas, graphs, and mathematical tables}, Dover, 9th ed., 1972.
\bibitem{Zhu2} B. Zhu, J. Cheng, J. Yuan, J.-Y Wang, L. Wu, and Y. Wang, ``A new asymptotic analysis technique for diversity receptions over correlated lognormal fading channels,'' \emph{IEEE Trans. Commun.}, vol. 66, no. 2, pp. 845--861, Feb. 2018.
\end{thebibliography}
\end{document}